\documentclass[prd,twocolumn,floats,superscriptaddress,eqsecnum,floatfix]{
revtex4}

\usepackage{amssymb,amsmath}
\usepackage{epsfig}
\usepackage{color}

\newcommand{\be}{\begin{equation}}
\newcommand{\ee}{\end{equation}}
\newcommand{\ba}{\begin{eqnarray}}
\newcommand{\ea}{\end{eqnarray}}

\newcommand{\const}{\mbox{const}}

\newcommand{\eq}[1]{Eq.(\ref{#1})}

\newcommand{\hh}{\, ,\hspace{0.8cm}}

\newcommand{\ins}[1]{{\mbox{\tiny #1}}}

\begin{document}

\title{Anomaly and the self-energy of electric charges}

\author{Valeri P. Frolov}%
\email[]{vfrolov@ualberta.ca}
\affiliation{Theoretical Physics Institute, Department of Physics,
University of Alberta,\\
Edmonton, Alberta, T6G 2E1, Canada
}
\author{Andrei Zelnikov}%
\email[]{zelnikov@ualberta.ca}
\affiliation{Theoretical Physics Institute, Department of Physics,
University of Alberta,\\
Edmonton, Alberta, T6G 2E1, Canada
}


\begin{abstract}
We study the self-energy of a charged particle located in a static
D-dimensional gravitational field. We show that the energy functional
for this problem is invariant under an infinite dimensional (gauge)
group of transformations parametrized by one scalar function of
$(D-1)$ variables. We demonstrate that the problem of the calculation
of the self-energy  for a pointlike charge  is equivalent to the
calculation of the fluctuations $\langle \psi^2\rangle$ for an
effective $(D-1)-$dimensional Euclidean quantum field theory. Using
point-splitting regularization we obtain an expression for the self-energy
and show that it possesses anomalies. Explicit calculation of
the self-energy and its anomaly is done for the higher dimensional
Majumdar-Papapetrou spacetimes.
\end{abstract}


\maketitle

\section{Introduction}

Recently it was demonstrated that the problem of the self-energy of a
pointlike scalar charge in a D-dimensional static gravitational field
can be  reduced to the problem of calculation of vacuum fluctuations
$\langle \varphi^2\rangle$ of a scalar field in the
effective $(D-1)-$Euclidean quantum field theory. This theory, besides
a dynamical field ${ \varphi}$ includes \mbox{$(D-1)-$dimensional} metric
$g_{ab}$ and a dilaton field $\alpha$, which is related to the $g_{tt}$
component of the metric of the original theory. Moreover, the energy,
which plays the role of the \mbox{$(D-1)-$dimensional} Euclidean action,
possesses the property of gauge invariance with respect to joint
transformation of ${ \varphi}$, $g_{ab}$, and $\alpha$. Standard
regularizations, required to make the self-energy finite, break this
invariance. As a result, the renormalized expression for the
self-energy and the mass shift for a scalar pointlike charge contains an
anomaly. This anomalous contribution vanishes for even D and is
nontrivial in odd dimensional spacetimes. This anomaly was calculated
and discussed in \cite{FrolovZelnikov:2012b}.

In this paper we demonstrate that a similar anomaly exists for the
self-energy of static pointlike electric charges in the higher
dimensional Maxwell theory. In a static \mbox{$D-$dimensional} spacetime, this
problem also can be reduced to the calculation of the renormalized
value of the fluctuations $\langle \psi^2\rangle$ for some effective
\mbox{$(D-1)-$dimensional} Euclidean quantum scalar field $\psi$. The
effective $(D-1)-$dimensional action for this theory again is gauge
invariant for special transformation of the field and background
variables. We calculate the corresponding anomaly, and apply the
obtained results for the calculation of the electromagnetic mass shift
in some special static spacetimes, containing black holes.

\section{self-energy of an electric charge in static
spacetimes}\label{self-energy}

Let us consider an electric charge $e$
in a static $D-$dimensional spacetime with the metric $g_{\mu\nu}$
\be\begin{split}\label{ds}
&ds^2=-\alpha^2 dt^2+g_{ab}\,dx^a dx^b \,.
\end{split}\ee
We assume that the spacetime is static $\partial_t\alpha=\partial_t g_{ab}=0$.
The action for the Maxwell field in higher dimensions is
\be\label{Maxwell}
I=-{1\over 16\pi}\int d^Dy\,\sqrt{-{\rm g}}\,F^{\mu\nu}F_{\mu\nu}
+\int d^Dy\,\sqrt{-{\rm g}}\, A_{\mu }\,J^{\mu }\,.
\ee
Here $F_{\mu \nu }\equiv\partial_{\mu} A_{\nu}-\partial_{\nu} A_{\mu}$ and
\be
{\rm g}=\det g_{\mu\nu}=-\alpha^2\,g\hh
g=\det g_{ab}\, .
\ee
The Greek indices $\alpha,\beta,\ldots$ mark spacetime coordinates, while Latin indices $a,b,\ldots$ correspond to spatial coordinates.

The field obeys the equation
\be\label{divF}
F^{\mu\epsilon}{}_{;\epsilon}=4\pi J^{\mu}\, .
\ee
The electromagnetic stress-energy tensor is
\begin{equation}
T_{\mu \nu }=\frac{1}{4\pi }\left( F_{\mu \alpha }\,F_{\nu
}{}^{\alpha}-\frac{1}{4}\,g_{\mu \nu }\,F_{\alpha \beta }\,F^{\alpha \beta
}\right)\,.
\end{equation}
For a static source $J^{\mu}=\delta^{\mu}_0 J^0$ the only nonvanishing
components of the Maxwell tensor are $F_{0 a}=-F_{a 0}$. In the Coulomb gauge the Maxwell tensor reads
$
F_{0 a}=-\partial_{a} A_{0}\,.
$
The nontrivial Maxwell equations
\be
F^{0\epsilon}{}_{;\epsilon}=4\pi J^{0}\,,
\ee
when rewritten in terms of the vector potential $A_{\alpha}=(A_0,A_a)$, are
\be\label{eqA0}
{1\over \sqrt{-\rm g}}\partial_{\alpha}\left(\sqrt{-\rm g}\,g^{00}g^{\alpha\beta}
\partial_{\beta}A_0\right)=-4\pi J^0\,.
\ee
Without loss of generality, one can put $A_{a}=0$ and fix the asymptotic value of the potential at spatial infinity $A_0|_{\infty}=0$.

The energy $E$ of a static configuration of fields is
\be\begin{split}\label{E}
E=-{1\over 8\pi}&\int_{\cal M} d^{D-1}x\,{ \sqrt{-\rm g}}\,F^{0a}F_{0a}\\
={1\over 8\pi}&\int
d^{D-1}x\,\sqrt{g}\,\alpha^{-1}\,g^{ab}\partial_{a}A_{0}\,\partial_{b}
A_{0}
\, .
\end{split}\ee
In the presence of black holes the integration is over the space ${\cal M}$ lying
outside the horizons.
By integration by parts we rewrite this spatial integral as the bulk integral and a set of
\mbox{$(D-2)-$dimensional} surface integrals over the black hole horizons
$\Sigma_{k}$ and boundary at spatial infinity $\Sigma_{\infty}$.
\be\begin{split}
E=-{1\over 2}&\int_{\cal M} d^{D-1}x\,{ \sqrt{-\rm g}}\,A_{0}J^{0}\\
-{1\over 8\pi}&\int_{\Sigma_{\infty}}d\sigma_a { \sqrt{-\rm g}}\,
g^{00}g^{ab} A_0\partial_b A_0\\
{1\over 8\pi}&\int_{\Sigma_{k}}d\sigma_a { \sqrt{-\rm g}}\,
g^{00}g^{ab} A_0\partial_b A_0
\, .
\end{split}\ee

We choose $A_0\big|_{\Sigma_{\infty}}=0$ at the boundary at infinity. The
boundary integral at infinity is proportional to the charge of the source and to
$A_0\big|_{\Sigma_{\infty}}$. Hence it vanishes.

The generic
property of black holes is that the vector potential $A_0$ on every horizon is constant.
Therefore the surface integrals over every horizon are proportional to these
constants and to the total flux across the
horizon of the electric field created by the source. Because the charge is
located outside the horizons this flux is identically zero. Thus all surface
integrals vanish. The bulk integral can be represented in terms of the static
Green function ${\cal G}_{00}$ of the Maxwell field
\be
A_0=4\pi \int_{\cal M}d^{D-1}x'\,{ \sqrt{-{\rm g}(x')}}\,{\cal
G}_{00}(x,x')J^{0}(x')
\ee
where ${\cal G}_{00}$ is the solution of the problem
\be\label{eqA0}
{1\over \sqrt{-\rm g}}\partial_{a}\left(\sqrt{-\rm g}\,g^{00}g^{ab}
\partial_{\beta}\right){\cal G}_{00}=-{\delta^{D-1}(x-x')\over \sqrt{-\rm g}}\,.
\ee
Eventually we get
\be\begin{split}\label{E1}
E=-2\pi\int dx^{D-1} d^{D-1}x'{ \sqrt{{-\rm g}(x)}}
{ \sqrt{-{\rm g}(x')}}&\\
\,J^{0}(x)\,{\cal G}_{00}(x,x')\,J^{0}(x')
\, .&
\end{split}\ee

It is convenient to introduce another field variable $\psi$ instead of
the electric potential
\be
A_0=-\alpha^{1/2}\,\psi\,.
\ee
Then we can rewrite our problem as that for the scalar field $\psi$ in
\mbox{$(D-1)-$dimensional} space and in the presence of the external dilaton
field $\alpha$.

The equation for the field $\psi$ can be derived from \eq{eqA0}
\be\begin{split}\label{eqpsi}
{\cal O}\,\psi&=-4\pi j\hh
{\cal O} \equiv (\triangle  +V)\,.
\end{split}\ee
Here,
\be
\triangle=g^{ab}\nabla_a\nabla_b\,
\ee
is the $(D-1)-$dimensional covariant Laplace operator,
$V$ is the potential, and $j$ is the effective scalar charge density
\be\begin{split}
V&=-{3\over 4}{(\nabla\alpha)^2\over
\alpha^2}+{\triangle\alpha\over 2\alpha}
\equiv -\alpha^{1/2}\triangle(\alpha^{-1/2})\,, \\
j&\equiv\alpha^{3/2} J^0
 \, .
\end{split}\ee
The field $\psi$ is chosen in such a way that the operator
$\cal O$ is self-adjoint in the space with the metric $g_{ab}$.

In terms of the field $\psi$ the energy \eq{E} takes the form
\be\begin{split}\label{E2}
E&={1\over 8\pi}\int
d^{D-1}x\,\sqrt{g}\,g^{ab}\left(\psi_{,a}+{\alpha_{,a}\over
2\alpha}\psi\right)\left(\psi_{,b}+{\alpha_{,b}
\over
2\alpha}\psi\right)
\, ,
\end{split}\ee
or, taking into account \eq{E1}, one can write
\be\begin{split}\label{E3}
E=-2\pi\int {  dx^{D-1} {dx'}^{D-1}}\,\sqrt{g(x)}\sqrt{
g(x')}&\\\,j(x)\,{\cal G}(x,x')\,j(x')
\, .&
\end{split}\ee
Here ${\cal G}(x,x')$ is the Green function, corresponding to the operator
${\cal O}=\triangle+V$,
\be\label{eqG}
(\triangle +V)\,{\cal G}(x,x')=-\delta^{D-1}(x,x')
\ee
The Green functions ${\cal G}$ and ${\cal G}_{00}$ are related to each other as follows
\be
{  {\cal G}_{00}(x,x')=-  \alpha^{1/2}(x)\,\alpha^{1/2}(x')\,{\cal G}(x,x')\,.}
\ee

\section{Symmetry property of the self-energy}

Consider the following transformations of the metric \eq{ds} and of the
field $\psi$
\be\begin{split}\label{trans}
&g_{ab}=\Omega^2 \bar{g}_{ab}\hh \alpha=\Omega^{n}\bar{\alpha}\hh
\psi=\Omega^{-n/2}\bar{\psi}\,,
\end{split}\ee
where
$
n\equiv D-3\,.
$
For these transformations one has
\be
A_0=\bar{A}_0\,.
\ee

From the point of view of a field theory on a \mbox{$(D-1)-$dimensional}
spatial
slice, \eq{trans} describe simultaneous conformal transformation of the metric
$g_{ab}$ and transformation of
the  dilaton field $\alpha$. Under these transformations the energy
functional \eq{E}
remains invariant.

The operator ${\cal O}$ in \eq{eqpsi} transforms homogeneously
\be\label{barO}
{\cal O}=\Omega^{-2-{n\over 2}}\,\bar{\cal O}\,\Omega^{n\over 2}\,,
\ee
That is
\be
(\triangle  +V)\,\psi = \Omega^{-2-{n\over 2}}(\bar{\triangle}
+\bar{V})\,\bar{\psi}\,,
\ee
It should be noted that these symmetry
transformations \eq{trans} differ from those for the self-energy of a scalar charge
\cite{FrolovZelnikov:2012b}. In the scalar case the transformation
of the dilaton field $\alpha$ has different power of the conformal
factor.

The energy $E$ \eq{E} is a functional of \mbox{$(D-1)-$dimensional} dynamical
field
$\psi$ and two external fields $g_{ab}$ and $\alpha$. The
transformations
\eq{trans} preserve the value of this functional. Our effective
$(D-1)-$dimensional Euclidean field theory happens to be invariant
under
infinite-dimensional group parametrized by one function $\Omega(x)$.
This property is similar to the conformal symmetry. It becomes the
conformal invariance of the theory only in four dimensions. Let us
note that the \eq{divF} for the Maxwell field $A_\mu$ in
spacetimes with the dimension $D>4$ is not conformally invariant.

\section{Classical anomaly}

The invariance with respect to the transformations \eq{trans} describes the
classical symmetry of the system. For a pointlike charge, the
classical functional \eq{E} diverges. The divergent part of the electromagnetic
energy of a charge can be recombined with the contribution of
nonelectromagnetic fields, which are responsible for the stability of the
charge and also contribute to its bare mass. After this renormalization one
obtains the finite total mass of the charge. But in a general case nonelectromagnetic fields do not respect the observed symmetry. This is the cause of an anomalous
contribution to the self-energy of charges in curved spacetimes. In the limit of
a pointlike charge, the details of the structure of the classical model of the
charge become unimportant and can be described by a universal function. The
regularization methods of quantum field theory provide us with the proper
tools to deal with the divergencies. In quantum field theory the fact that
renormalization procedure breaks
some symmetries of the classical theory is the cause of appearance of conformal,
chiral and other anomalies. In our case  the same arguments are applicable to
the renormalized self-energy of classical sources. For the same reason their
self-energy acquires anomalous terms. All
traditional methods of UV regularization like point-splitting, zeta-function
and dimensional regularizations, proper time cutoff,  Pauli-Villars and other
approaches are applicable to the calculation of the self-energy. For our problem
the most natural choice is to use the point-splitting regularization.

The self-energy of a scalar charge has been studied in our previous
papers \cite{FrolovZelnikov:2012a,FrolovZelnikov:2012b}. We have shown that in a generic case it
can be written as a sum of the self-energy in some reference spacetime and the
anomalous term. The anomaly proves to vanish in even-dimensional spacetimes,
while is nontrivial in odd-dimensional spacetimes.

The case of an electric charge can be treated along the same lines, in spite of the
fact that the symmetry transformations \eq{trans} are different from those of
the scalar sources. The renormalized self-energy \eq{E3} of a
pointlike charge $e$ described by the current
\be
J^0=e\,\alpha^{-1}\delta^{n+2}(x,x')
\ee
takes the form
\footnote{Note that here we use Gaussian units. For calculation of the vacuum fluctuations in quantum field theory it is more traditional to use Heaviside-Lorentz units which differ from the Gaussian units by numerical factors. In Heaviside-Lorentz units the first term in the Maxwell action \eq{Maxwell} has the coefficient $1/4$ instead of $1/16\pi$. The substitution $A_{\mu}\rightarrow A_{\mu}\sqrt{4\pi}$ and  $J^{\mu}\rightarrow J^{\mu}/ \sqrt{4\pi}$ transforms Gaussian form of the equations to the Heaviside-Lorentz one. As the consequence $e^2\rightarrow e^2/(4\pi)$ and the self-energy in Heaviside-Lorentz units acquires an additional $1/(4\pi)$ factor. The mass shift of a scalar charge $\Delta m=q^2{\cal R}/576\pi^2$, which is given by Eq.(5.13) of the paper \cite{FrolovZelnikov:2012a},  corresponds to Heaviside-Lorentz units. In Gaussian units it reads $\Delta m=q^2{\cal R}/144\pi$.
}
\be
E_\ins{ren}\equiv\alpha\,\Delta m \,,
\ee
\be\label{E_ren}
{ \Delta m=2\pi e^2 {\cal G}_\ins{reg}(x,x) \, .}
\ee

Here ${\cal G}_\ins{reg}(x,x)$ is the coincidence limit $x'\rightarrow x$ of
the regularized Green function
\be
{\cal G}_\ins{reg}(x,x')=
{\cal G}(x,x')-{\cal
G}_\ins{div}(x,x')\, .
\ee
The Green function
${\cal G}$ corresponds to the operator \eq{eqpsi}.
Thus, in order to find out the self-energy of an electric charge one has to know
the regularized Euclidean Green functions ${\cal G}_{00}{}_\ins{reg}(x,x)$ or
${\cal G}_\ins{reg}(x,x)$. The latter one
in the limit of coincident points is exactly the
$\langle\psi^2\rangle_\ins{ren}$ of the scalar field $\psi$. In other words
technically the problem of calculation of $\Delta m$ is formally equivalent to
computation of the quantum vacuum average value of
$\langle\psi^2\rangle_\ins{ren}$ in $(D-1)-$dimensional space.

Similar to the scalar case,  \cite{FrolovZelnikov:2012b} the Green
function \eq{eqG} transforms as follows
\be
{\cal G}(x,x')=\Omega^{-{n\over 2}}(x)\,\bar{\cal G}(x,x')\,\Omega^{-{n\over
2}}(x')\,.
\ee
Formally the nonrenormalized value of
$\langle\psi^2\rangle$ transforms homogeneously
\be
\langle\psi^2(x)\rangle=\Omega^{-n}(x)\,\langle\bar{\psi}^2(x)\rangle\,.
\ee
Thus the combination
\be
g^{n\over 2(n+2)}\langle\psi^2\rangle
\ee
is formally invariant under the transformations \eq{trans}.
The regularization procedure breaks this classical symmetry and
$
g^{n\over 2(n+2)}\langle\psi^2\rangle_\ins{ren}
$
is not invariant anymore.
However, one can find such finite term ${\cal A}(x)$ that makes the
combination
\be\label{phi2}
g^{n\over 2(n+2)}\left(\langle\psi^2\rangle_\ins{ren}+{\cal A}\right)
\ee
invariant.

Following exactly to the lines of the paper \cite{FrolovZelnikov:2012b} one can
show that
\be
\langle\psi^2\rangle_\ins{ren}=\Omega^{-n}\,\langle\bar{\psi}
^2\rangle_\ins {ren}-{\cal B}\,,
\ee
where
\be\label{BA}
{\cal B}(x)={\cal A}(x)-\Omega^{-n}\bar{{\cal A}}(x)
\ee
is defined as
\be
{\cal B}(x)=\lim_{x'\rightarrow x}\left[ {\cal G}_\ins{div}(x,x')-
{\bar{\cal G}_\ins{div}(x,x')\over\Omega^{{n/
2}}(x)\,\Omega^{{n/
2}}(x')}\right]\,.
\ee
In order to find the explicit form of this finite anomalous term ${\cal B}$ one can
apply
the Hadamard representation of the Green functions. The divergent part
of the Green function \cite{FrolovZelnikov:2012a} reads
\be\begin{split}\label{calGdiva}
{\cal G}_\ins{div}(x,x')
&={\Delta^{1/2}(x,x'){1\over (2\pi )^{{n\over 2}+1}}}\,\\
&\times\sum_{k=0}^{[n/2]}
{\Gamma\left({{n\over 2}-k}\right)\over
2^{k+1}\sigma^{{n\over2}-k}}{a}_k(x,x') \,.
\end{split}\ee
For even  $n$ the last term $(k=n/2)$ in the sum is to be replaced by
\be\begin{split}
{\Gamma\left({{n\over 2}-k}\right)\over
2^{k+1}{\sigma}^{{n\over
2}-k}}\,&
{a}_k(x,x')\Big|_{k=n/2}\\
&\rightarrow
-{\ln{\sigma}(x,x')+\gamma-\ln
2\over 2^{{n\over
2}+1}}\,{a}_{n/2}(x,x') .
\end{split}\ee
Here ${a}_k(x,x')$  are the Schwinger--DeWitt coefficients
for the operator $O$. The world function
${\sigma}(x,x')$ and Van Vleck--Morette determinant
$\Delta(x,x')$ are defined on the $(n+2)-$dimensional space with the metric
$g_{ab}$.

Thus, the self-energy of an electric charge takes the form $E_\ins{ren}\equiv\alpha\,\Delta m$, where
\be\begin{split}\label{Delta_m}
\Delta m=&{ \,2\pi e^2 \,\langle\psi^2\rangle_\ins{ren}}\\
=&{ \,2\pi e^2  \left[\Omega^{-n}\,\langle\bar{\psi}
^2\rangle_\ins {ren}-{\cal B}\right]
} \, .
\end{split}\ee
The first term in brackets describes the proper mass calculated in a reference spacetime $\bar{g}_{\mu\nu}$. This term takes into account the dependence of the self-energy on boundary conditions and other IR properties of the electromagnetic field in curved spacetime. The second term comes out of the anomaly in question and reflects UV behavior of the system.

There are physically interesting spacetimes where the self-energy can be computed exactly.
At any rate the anomalous contribution is local and it can be computed exactly in any static spacetime. In several special cases the reference space can be chosen in such a way that the calculation of vacuum fluctuations $\langle\bar{\psi}^2\rangle_\ins {ren}$ becomes simple or exactly solvable.

The scalar and electric charges near the extremal charged black hole in higher dimensions happen to belong to this class of exactly solvable models.

\section{Higher dimensional Majumdar-Papapetrou spacetimes}

In the case of the higher dimensional Majumdar-Papapetrou spacetimes,
\cite{Myers:1986}
\be\begin{split}\label{MP}
ds^2&=-U^{-2}\,dt^2+U^{2/n}\delta_{ab}\,dx^a dx^b\, ,\\
A^\ins{MP}_{\mu}&=\sqrt{{n+1\over 2n}}\,U^{-1}\,\delta^0_{\mu}\,.
\end{split}\ee
the transformations \eq{trans} with
\be
\Omega=U^{1/n}
\ee
lead to the metric $\bar{g}_{\mu\nu}$ with
\be
\bar{\alpha}=U^{-2}\hh\bar{g}_{ab}=\delta_{ab}\,.
\ee
Therefore the potential for the field $\bar{\psi}$ in the
reference background metric $\bar{g}_{\mu\nu}$ becomes
\be
\bar{V}=-U^{-1}\vartriangle\! U
\ee
where
\be
\vartriangle=\delta^{ab}\partial_a\partial_b
\ee
is the Laplace operator corresponding to the flat spatial metric
$\bar{g}_{ab}=\delta_{ab}$.

The function $U$ describing the Majumdar-Papapetrou metrics \eq{MP}
satisfies the equation
\be\label{lapU}
\vartriangle\! U =-{4\pi^{1+{n\over 2}}\over \Gamma\left({n\over
2}\right)}\sum_k
M_k\,\delta^{n+2}(\boldsymbol{x}-\boldsymbol{x}_k)\, .
\ee
The explicit solution of the \eq{lapU} reads
\be\label{U}
U=1+\sum_k {M_k\over \rho_k^n}\hh
\rho_k=|\boldsymbol{x}-\boldsymbol{x}_k|\, .
\ee
Here $\boldsymbol{x}\equiv x^a$ and
$|\boldsymbol{x}|^2\equiv\delta_{ab}\,x^ax^b$.
The index $k=(1,\dots,N)$ enumerates the extremal black holes and
$x^a_k$ marks the spatial position of the $k$-th black
hole.

Thus for the Majumdar-Papapetrou metrics, the potential $\bar{V}$
vanishes everywhere, except the location of the horizons, where
$U\rightarrow\infty$. On the horizon, $\bar{V}$ formally
vanishes but, because the field $\psi$ may diverge there,
one has to be quite accurate in dealing with distributions
\cite{FrolovZelnikov:2012}. In fact, the $\delta-$functions do contribute to
the potential $\bar{A}_0$ and the self-energy of electric charges.

The solution for the static Green function has been obtained in
\cite{FrolovZelnikov:2012}
\be\label{G00}
{\cal G}_{00}(x,x')=-{\Gamma\left({n\over
2}\right)\over
4\pi^{1+{n\over 2}}}
{1\over U(x)U(x')}\left[{1\over R^n}+\sum_k
{M_k\over
\rho_k^n {\rho'_k}^n}\right]\,,
\ee
\be\label{R}
R=|\boldsymbol{x}-\boldsymbol{x}'|\,.
\ee
The Green function ${\cal G}$ for the scalar field $\psi$ obeying \eq{eqG}
is
\be\label{calG}
{  {\cal G}(x,x')={\Gamma\left({n\over
2}\right)\over
4\pi^{1+{n\over 2}}\sqrt{UU'}}
\left[{1\over R^n}+\sum_k
{M_k\over \rho_k^n {\rho'_k}^n}\right]\,.
}
\ee
In any reference spacetime with the metric $\bar{g}_{\alpha\beta}$ described by \eq{trans} the corresponding Green function
\be\label{barG}
{ \bar{\cal G}(x,x')={\cal G}(x,x')\,\sqrt{U(x)U(x')}\,.}
\ee

In the case of Majumdar-Papapetrou metrics the reference spacetime can be chosen
such that $\bar{g}_{ab}$ is a flat metric. In this case
the vacuum divergent part of the `reference' Green function is evidently described by the
first term in brackets in \eq{calG}
\be\label{calGdiv}
{ \bar{\cal G}_{\ins{div}}(x,x')={\Gamma\left({n\over
2}\right)\over
4\pi^{1+{n\over 2}}}
{1\over R^n}\,.
}
\ee
Thus, the exact formula for the regularized Green function in higher
dimensional Majumdar-Papapetrou spacetimes is
\be\label{calGreg}
{ \bar{\cal G}_{\ins{reg}}(x,x')={\Gamma\left({n\over
2}\right)\over
4\pi^{1+{n\over 2}}}
\sum_k
{M_k\over \rho_k^n {\rho'_k}^n}\,.
}
\ee

Substitution of this expression to \eq{Delta_m} gives
\be\label{Delta_m1}
\Delta m=e^2\,{\Gamma\left({n\over 2}\right)\over 2\pi^{n\over 2}} U^{-1}
\sum_k
{M_k\over \rho_k^{2n}}+2\pi e^2\,{\cal B}\,,
\ee

In even-dimensional spacetimes, the anomaly ${\cal B}=0$.
In the case of a charge near a single extremal Reissner-Nordstr\"{o}m black hole
of mass $M=Q$ in four dimensions, the \eq{Delta_m1} reproduces a well known result \cite{FrolovZelnikov:1982,Lohiya:1982}
\be
E_\ins{ren}=e^2{M\over 2(\rho+M)^2}=e^2{M\over 2r^2}\,,
\ee
where $r=\rho+M$ is the Schwarzschild radial coordinate.

In odd dimensions the anomaly contribution enters the answer.
In five dimensions we can use the result of our previous paper \cite{FrolovZelnikov:2012b}, where we have derived that for arbitrary static spacetimes
\be\begin{split}
{\cal A}(x)&={1\over 288\pi^2}{\cal R}-{1\over 64\pi^2}\ln(g)\,a_1(x)\,,\\
a_1(x)&={1\over 6}{\cal R}+V \,.
\end{split}\ee
In our particular case of the Majumdar-Papapetrou spacetimes,
the DeWitt coefficient $a_1=0$ and applying \eq{BA} one gets
\be
{\cal B}={1\over 288\pi^2}{\cal R}\,,
\ee
where
\be
{\cal R}={3\over 2}U^{-3}\delta^{ab}\partial_a U \partial_b U-3 U^{-2}\vartriangle\! U
\ee
is the scalar curvature of the four-dimensional spatial slice $t=\const$ which is described by the metric \mbox{$g_{ab}=U\,\delta_{ab}$}. Note that on the Majumdar-Papapetrou spacetimes, the term $U^{-2}\vartriangle\! U$ vanishes both outside and at the horizons.

Thus, for $D=5$ we get $\Delta m=UE_\ins{ren}$ and
\be
E_\ins{ren}={e^2\over 2\pi}{1\over U^2}\sum_k
{M_k\over \rho_k^{4}}+{e^2\over 144\pi}{1\over U}{\cal R}\,,
\ee
where $U$ is given by \eq{U} for $n=2$. For a single extremal
Reissner-Nordstr\"{o}m black hole the self-energy is
\be
E_\ins{ren}={e^2\over 2\pi}
{M\over (\rho^2+M)^2}+{e^2\over
24\pi}{\rho^2M^2\over(\rho^2+M)^4}\,,
\ee
In terms of the Schwarzschild radial coordinate $r$, which is given by
$r^2=\rho^2+M$, the self-energy of the electric
charge $e$ near the five-dimensional extremal Reissner-Nordstr\"{o}m black hole has the form
\be
E_\ins{ren}={e^2\over 2\pi}
{M\over r^4}+{e^2\over 24\pi}{(r^2-M)\,M^2\over r^8}\,.
\ee

\section{Conclusions}

In this paper we demonstrated that renormalized shift of the self-mass
for a pointlike electric charge in a static odd dimensional spacetime
contains an anomaly contribution. This is a result of the gauge
invariance of the corresponding $(D-1)-$dimensional Euclidean theory,
which is broken by a covariant regularization. There exists a certain
similarity of this property of the electromagnetic self-energy and the
property of the self-energy for a scalar massless field. However, the
corresponding form of gauge transformations preserving the form of the
energy action is different for these two cases. It is quite
interesting that a solution of the rather old classical problem of the
self-energy in both cases can be reduced to the problem of
Euclidean quantum field theory in the space of the codimension one. This
makes it possible to use well developed tools of quantum theory to
perform the classical calculations. In particular, the results of
\cite{Moretti:1999} allow one to expect that the renormalized value of the
mass-shift and the calculated anomalies do not depend of the details
of the regularization scheme. Speaking in terms of the original
$D-$dimesional classical theory, the above described gauge
transformations relate spacetimes with nontrivial transformation of
the $D-$dimensional metric. In special cases one can use this
transformation in order to simplify the problem. We demonstrated that
for the wide class of Majumdar-Papapetrou spacetimes this allows one
to obtain an explicit answer. One can also try to apply the gauge
induced transformation in order to find a spacetime where appropriate
approximation can be used. Using the known anomaly one can obtain a
corresponding approximation for the mass-shift in the original,
physically interesting case. An interesting question is whether the
developed approach can be generalized for the calculation of the
gravitational self-energy
in static spacetimes. Another interesting question is: whether it is
possible to generalize this approach to stationary spacetimes. In more general terms,
it is interesting whether quantum tools can be used for solving other
classical problems, for example, calculation of the self-energy for
special distributed configuration of charges and/or self-energy of
dipole (multipole) charge configurations.

\acknowledgments

This work was partly supported  by  the Natural Sciences and Engineering
Research Council of Canada. The authors are also grateful to the
Killam Trust for its financial support.

\end{document}